# Promoting Reliable Knowledge about Climate Change: A Systematic Review of Effective Measures to Resist Manipulation on Social Media

Aliaksandr Herasimenka[1,2*], Xianlingchen Wang[1,3], Ralph Schroeder[1]

[1] Oxford Internet Institute, University of Oxford, Oxford, United Kingdom

[2] Department of Communication and Media, University of Liverpool, Liverpool, United Kingdom

[3] Department of Population Health, University of Oxford, Oxford, United Kingdom

[*] Corresponding author can be contacted at a.herasimenka@liverpool.ac.uk

# Promoting Reliable Knowledge about Climate Change: A Systematic Review of Effective Measures to Resist Manipulation on Social Media

## Abstract

We present a systematic review of peer-reviewed research into ways to mitigate manipulative information about climate change on social media. Such information may include disinformation, harmful influence campaigns, or the unintentional spread of misleading information. We find that commonly recommended approaches to addressing manipulation about climate change include corrective information sharing and education campaigns targeting media literacy. However, most relevant research fails to test the approaches and interventions it proposes. We locate research gaps that include the lack of attention to large commercial and political entities involved in generating and disseminating manipulation, video- and image-focused platforms, and computational methods to collect and analyze data. Evidence drawn from many studies demonstrates an emerging consensus about policies required to promote reliable knowledge about climate change and resist manipulation.

# Introduction

Despite solid scientific consensus about the anthropogenic roots and consequences of climate change, false, inaccurate, or misleading online content often confuses the public, undermining support for urgent mitigating policies (1). This is important because belief in climate change is more easily weakened than strengthened (2). Terms used to identify and describe such content include 'manipulation,' 'misinformation,' 'disinformation,' 'fake news,' or 'propaganda.' Such manipulative content often spreads on social media platforms (3,4), which are a particularly important source of information about climate change for at least a third of the populations of some of the largest economies, compared with other channels (5).

This review discusses this unreliable and misleading content about climate change under the wider umbrella concept of 'manipulative information about climate change', which encompasses phenomena ranging from misleading information to attempts at deliberately distorting the information environment.

It has been argued that the positive effects of reliable information can far outweigh the effects of manipulation (6). Plenty of studies have analyzed who spreads manipulated information online, how it spreads, and its consequences, including experiments that test solutions to this problem. However, we also know that the climate change debate is *particularly* susceptible to the diffusion of such information (7). This might be linked to the prominence of this issue on media agendas, but the major economic and political interests at stake also play a role.

To progress our understanding of these topics, it is important to synthesize the research into the sources, channels, and other factors of manipulation spread. However, a systematic review that would synthesize research into manipulative information about climate change on social media has been missing (4), as a broader literature on the phenomenon of climate denial has started to be systematically analyzed only recently (8). At the same time, 'the first systematic and critical review of the literature on social media and climate change' was published only in 2019 (9).

Our systematic approach has several advantages compared to recent reviews of communication about climate change on social media (4,9). It expands and improves on the existing approaches taken in the early reviews of climate change manipulation by searching multiple databases of literature, undertaking a search with multiple terms, detailing our methodology, following the recommendation of the Preferred Reporting Items for Systematic Reviews and Meta-Analyses (PRISMA) (10). Our review also focuses on perhaps the most problematic environment where current climate change manipulation originates, that is, social media platforms.

We found that commonly recommended approaches to addressing manipulation about climate change include corrective information sharing and education campaigns targeting media literacy. However, most publications did not test the approaches and interventions they proposed. Research gaps that we located include the lack of attention to large commercial and political entities involved in generating and disseminating manipulation, video- and image-focused platforms, and computational methods to collect and analyze data.

## Misinformation and Manipulation

It is necessary to examine what is meant by 'manipulation' about climate change on social media. The literature uses multiple terms to describe the attempts to discredit climate science and sow confusion in the debate on climate change by casting doubt on well-supported evidence and promoting often fallacious interpretations of observations (4). Treen and colleagues collected several such terms, including 'skeptical discourse,' 'contrarian information,' or 'denial campaigns.' The most popular of these terms relates to the concept of 'climate skepticism.' Authors often apply this concept to mean those who doubt climate change or reject mainstream climate science, instead of its original meaning referring to judging 'the validity of a claim based on objective empirical evidence,' an integral part of the scientific method (4,11). Such indiscriminate use of terms may appear to have negative consequences as it polarizes 'views on climate change […] and do little to advance the unsteady relationship among climate science, society, and policy' (12).

At the same time, our field has experienced a reframing of the debates about communicating on social media with increased interest in such phenomena as misinformation and disinformation (13). This is also the case of the topic of climate change (4). The critical difference between 'misinformation' and 'disinformation' is the intention to mislead: while the former refers to information that may or may not inadvertently mislead, the former is intentional (13). Hence, reserving to one of these terms is too restrictive as the intention to mislead is often hard to determine.

Our study casts the distinction between misinformation, disinformation, and climate denialism in a new light. We do not argue that climate change denialism, or climate skepticism in its un-scientific meaning, has been studied insufficiently. However, we argue that the range of this phenomenon requires a broader focus that would encompass both intentional and unintentional attempts to influence the discussion of climate change through reliance on false information. We use the term 'manipulation' as an umbrella concept that is originally defined as 'influencing someone's beliefs, attitudes, or preferences in ways that fall short of what an empathetic observer would deem normatively appropriate in context' (14). This concept is appropriate for our aim as it does not imply intentionality like disinformation, does not confuse legitimate scientific judgment and denialism like climate skepticism, but helps to focus on manipulation sources with an economic, political, or social interest in spreading perceptions of climate change that are normatively inappropriate in context.

## Previous Reviews and Research Questions

Previous research on climate change is extensive, and there are good overviews that address social science research on the topic generally (e.g., Klinenberg, Araos and Koslov (15)). There is also a large body of research about the public understanding of science (16) and recent meta-analysis studies into the perceptions of climate information that are primarily informed by psychology (1,2,17). Some earlier work on communicating climate change was about media reporting (18), and more recent reviews focused on communicating 'climate effectively' (19).

A few recent reviews also analyzed how the use of the internet and social media affect public understanding of and engagement with climate change information. For example, Schäfer (20) and Pearce and colleagues (9) reviewed the literature on social media and climate change but mentioned misinformation only briefly. The former noted that 'although (or because) many stakeholders participate online, this does not lead to robust scientific information or better debates' (21).

Treen and colleagues (4) conducted an early review of climate change and sources of misinformation, including and going beyond social media. They found that most studies agreed that climate change misinformation has confused the public, led to political inaction, and stalled support for or led to the rejection of mitigation policies. Their research also has shown how even a small amount of misleading climate information, such as a few statistics, is effective in lowering people's acceptance of climate change. Psychology-informed meta-analyses found that values, ideologies, worldviews, and political orientation are the strongest predictors of belief in climate change (17), more than, for example, age or education (1).

These previous literature reviews inform our research questions, though relatively few of these reviews followed the PRISMA guidelines. First, the review studies often contextualize and differentiate between the approaches to examining climate communication by analyzing prevailing methodologies and the state of the art of the discipline. This helps to identify existing research gaps and emerging trends (22). For example, previous reviews have highlighted the Anglo-American focus of most studies in the area (8).

Second, there has been extensive research on communicating and perceiving climate change information, but this needs to be updated to consider how the public has turned to social media for information where misinformation might flourish. Literature in the broader field of social media misinformation studies considers three key questions: who spreads misinformation, what channels are used, and what can be done (3,23).

These are questions that some literature reviews also partially focus on. For example, Björnberg and colleagues (8) reviewed climate denialism beyond social media platforms and identified six types of sources of denying narratives: denying scientists, governments, media, political and religious organizations, industry, and the public. Treen and colleagues (4) emphasize the role of 'corporate and philanthropic actors with a vested interest [that] provide funding to […] produce climate change misinformation.' Overall, we now know that many entities have malicious aims in their climate communication, but the identification of these sources of manipulation has been rather rare in the reviews that specifically focus on social media platforms.

Third, the 'channels of spread' question is also at the center of some previous reviews. However, one of the most recent reviews agrees that 'there has been little research specifically into the diffusion of climate change misinformation' (4). Even more recent reviews in adjacent fields surprisingly give relatively little prominence to social media as a channel of misinformation communication (24). As with much research on digital communication, the vast bulk of earlier work focuses on Twitter—even though other social media are equally, if not more, important in climate change communication (9,25). The factors that exacerbate the spread of manipulative information are given more prominence, mainly thanks to the efforts of psychologists focusing on different misinformation domains. This research mainly focuses on personal traits, pre-existing beliefs, and ideology.

Finally, the intervention strategies reviewed often focus on broader solutions. Yet few, if any, studies have systematized interventions concerning the problem of manipulation or its manifestations on social media specifically. Björnberg and colleagues (8) name five 'strategies against denialism': the need for change, context-dependent strategies, communication strategies, education, and changing the focus of scientists. Developing critical thinking among the population seems to be the most commonly agreed upon idea to address this (4). Though, again, these findings are mostly based on data from North America. Several scholars advanced the general understanding of some appropriate measures that should be taken. These measures range from 'consensus messaging' (19) or 'morality' messaging (2) to targeting people's 'psychological factors' (17). However, there is a substantial bias towards social psychology-oriented solutions and relatively little discussion of what can be done on the policy and platform governance levels, such as how to communicate relevant messages. Consequently, given widespread and continuing manipulation campaigns online, how exactly the promotion of such climate change consensus can be achieved remains less clear.

In our systematic review, we go beyond previous literature review studies, systematic and not systematic, to identify how the research has progressed over time and the current methodologies that dominate the field. We also advance the field by expanding the focus across three key questions of social media misinformation studies, with the following research questions:

1. How has research about the dissemination of manipulation about climate change on social media progressed over time, and what methodologies has it adopted?
2. What sources and channels involved in disseminating manipulation about climate change does the literature focus on?
3. What factors exacerbating this dissemination does the literature examine?
4. What interventions have been proposed and shown to prevent or mitigate the impact of such manipulative information on social media?

**Methods**

Our methodological protocol followed the 2021 PRISMA guidelines (Supplementary Table S4). We searched for peer-reviewed articles, books, book chapters, and conference proceedings in two academic databases—Web of Science and SCOPUS ($N_{SCOPUS}$ = 236, $N_{WOS}$ = 142), which contained 128 duplicates. These databases offered a valid instrument for evaluating scholarly publications in social science (26) and have been used in past systematic reviews that asked similar questions (27,28). We searched for synonyms of 'manipulation,' in publication's abstracts, titles, and keywords, including any publication that used the terms 'misinformation,' 'disinformation,' 'propaganda,' 'fake news,' 'rumor.' We also searched for the term 'skeptical', which is commonly applied when discussing information about climate change in similar studies (4). See Supplementary Note SN1 for the complete search string.

We tested several search term combinations, but the one chosen ensured the highest recall. After testing, we deliberately refrained from searching for specific platform names, such as Twitter and YouTube, as we found that it skewed the search results by systematically overlooking lesser-known platforms. We also decided not to consider gray literature, such as non-peer-reviewed conference abstracts or presentations, because articles that have been through rigorous review are more likely to use complete methodology information, more refined analysis, and provide more transportable statistics.

Having searched the databases, we performed a supplementary check for any missed but relevant references, following an approach adopted by similar studies (29). For this, we consulted the reference lists of two review articles, Rode et al. (2021) and Treen et al. (2020) and used Google Scholar search. This added 43 additional publications and left us with 294 publications in total, as presented in a PRISMA flowchart (Fig 1).

Fig 1. Flow of publications through different stages of the systematic review

Supplementary Table S1 presents out codebook that we developed to assess the eligibility of the collected publications. This included publication day, the language of a publication, a focus on manipulation, platforms, and climate change (Supplementary Note SN2 defines each eligibility criterion). Two authors and a research assistant read the titles, abstracts, and keywords of the 293 publications to affirm their eligibility. To achieve an optimum level of reliability, the three coders ran a pilot test in pairs. Intercoder reliability for eligibility criteria based on a random sample of 146 publications showed an agreement above 0.6 (see Supplementary Table S3 for Krippendorff's α measures). Disagreements were discussed in the group and resolved (22). Some publications contained all search terms but did not consider the manipulation of climate change information as an object of study; they were not empirical or did not focus on social media platforms. In total, 38 publications met our eligibility criteria, all of which could be retrieved fully and thus were selected for the main stage of systematic analysis.

Our codebook for the full publication coding was drawn from previous studies of information manipulation on social media (30,31). We conducted a pilot coding exercise: the three coders compared their coding experiences and adopted the final coding template. All the publications were coded twice. The coders compared their article coding and resolved any discrepancies through conversation.

To address the first research question concerning the progress over time of research on the spread of climate change misinformation on social media, we turned to automatic analysis tools, including *bibliometrix* (32) and a custom script for R. We then summarized the dataset with manual coding, identifying the methods used in the publications. If a publication relied on several methods, both of them were coded. We then moved to research questions two, three, and four, which focus on sources, channels, and factors that facilitate the spread of climate change manipulation on social media, as well as the interventions proposed.

# Results

We have grouped the results into four sets based on the four research questions: the evolution and state of the art in research into manipulative information about climate change spreading on social media; the sources and specific communication channels involved in disseminating such information; factors exacerbating this dissemination; and proposed interventions to address this problem.

## State of the Art

Our review confirmed a trend towards an increase in interest in the topic of manipulation of climate change information online over the past years. Since 2015, the number of publications focusing on this issue has increased approximately five-fold from two to eleven, with especially noticeable growth since 2021 (Fig 2).

Fig 2. Temporal trends in eligible publications
Note: *n*=38 publications.

However, the geographic focus of this rapidly growing body of research remained highly Western-centric. In particular, our (English-language) literature sample is heavily dominated by authors based in the United States and the United Kingdom (Fig 3), which will have influenced the focus of the publications. Confirming and extending the findings of previous reviews in a systematic manner (4), we found that a lot of research focuses on the ideological debates in the US. Some studies focused on how climate change 'skeptics' manipulate public deliberations about the issue on platforms. For instance, Moernaut et al. (33) studied how ideological polarization influences people's belief in climate change conspiracies. They found that one side used problematic discursive strategies to delegitimize the other side 'as irrational, immoral or unnatural.' This helped to aggravate the polarization between climate change 'believers' and 'skeptics.' In addition, a few publications examine how large corporations fund controversial public relations and communication campaigns covering climate change.

Fig 3. Country of affiliation for the first authors of the eligible publications
Note: $n$=38 publications.

Content analysis was found to be the most common method used to study manipulative information about climate change on social media, with relatively few eligible studies conducting experiments or surveys (Fig 4). This means that most publications that proposed a policy or a solution to the problem of the spread of manipulative information about climate change did not test these interventions. Fig 7 shows the methods used to study the types of variables (we discuss these variables below). This shows that the research into the manipulation of climate change information could benefit from a more intensive application of computational content analysis and network analysis. We also found that the articles we analyzed were published in a diversity of venues, and none of them dominates the sample.

Fig 4. Methods used by the analyzed publications
Note: $n$=38 publications.

**Sources and Channels of Manipulation about Climate**

We found that some crucial sources of manipulative information about climate change are understudied by researchers focusing on social media. Fig 5 depicts the sources of manipulation named in the publications. Platform audiences sharing manipulative information about climate change, intentionally or not, were the major focus of these publications. At the same time, very few studies discussed such important sources as industry or governments that have previously been shown to be behind major campaigns covertly or openly propagating climate denial. For example, Supran and Oreskes found that ExxonMobil's public climate change messaging mimicked "tobacco industry propaganda" (34). See Table 1 for an overview of common manipulative narratives spread by different sources.

Fig 5. Sources of manipulation studied

Note: *n*=26 publications that identified or proposed the source of manipulation. Communities refer to defined or organized units on platform, such as groups or pages, while audiences refer to broader and undefined cohorts of users.

Table 1. Example of climate change manipulative narratives

| Source | Narratives |
| --- | --- |
| Social media platform users | |
| Twitter communities | global warming is a myth; |
| | climate activism is 'ecofascism'; |
| | climate change is not man-made; |
| | climate science is a conspiracy favoring growth of government; |
| | renewable energy as a misuse of taxpayers' money and an inappropriate manipulation of the market; |
| | climate science is merely sheep's clothing for the wolf of taxation; |
| | tens of thousands of commercial airliners a day are deliberately spraying some kind of mixture of toxic chemicals which influences environment; |
| | climate change is just a natural cycle |
| Facebook communities/friends | global warming is a myth; |
| | climate scientists cherry-pick their data |
| Instagram accounts | $CO_2$ is plant food and good for plants |
| Other Sources | |
| US junk news outlets, e.g. Breitbart, InfoWars, and Natural News | global warming is a myth; global warming causes global cooling |
| Corporate and philanthropic entities, political and religious organizations | global warming is a myth; climate alarmism |
| Contrarian/'Fake' scientists (e.g. Blair Fine) | 'Global warming is a hoax' |
| Google, Bing, and Yahoo search webpages | climate change's effects on agriculture positively |
| YouTube content by professional media organizations, e.g. The Weather Channel, The Onion, ABC News' Good Morning America | 'Scientists, the public, celebrities, and the media have accepted the conspiracy that climate change is manufactured' |

We also found that the most common channel for the dissemination of manipulated information examined in the publications was text-based social media posts appearing on older types of platforms—Twitter and Facebook. Visuals-focused channels like Instagram, TikTok, and YouTube received less attention—despite these latter platforms occupying a prominent place in the media diet of audiences (5). That said, about a third of the analyzed publications did not specify the channel they examined.

**Factors Exacerbating Manipulation Dissemination**

Our analysis shows that the literature has emphasized the personal traits, pre-existing beliefs, and ideology of consumers of information as variables that are more likely to exacerbate the diffusion of manipulative information about climate change. Hence, the literature focuses on what Chadwick and Stanyer (13) describe as the attributes and actions of the deceived and the role of beliefs in consequential attitudes and behavior. In particular, the publications we analyzed show that right-wing (35) or radical political ideology (36,37), trust in science (38), beliefs in a conspiracy (39), or inclination towards specific emotional reactions (40,41) can prompt individuals to share or engage with manipulated information about climate change. The scope of the analyzed literature is quite broad, which meant we had to group variables into wider categories for a summary (Fig 6).

Fig 6. Variables exacerbating the diffusion of manipulative information about climate change

Note: *n*=21; we excluded eligible studies with no identifiable variables.

Content exposure also featured high on the list of exacerbating variables. This refers to users' exposure to climate change content from *all* types of sources, including right-wing or fact-checking organizations, activists, or ordinary individuals discussing a climate disaster (36,42,43). However, we know less about content exposure effects as the studies tended to focus instead on a broader range of methods to study traits, beliefs, or ideology (Fig 7).

Fig 7. Distribution of variables exacerbating the diffusion of manipulative information about climate change over combinations of methods

Notes: *n*=21; we excluded eligible studies with no identifiable variables; if one publication reported several variables or used several methods, we counted them separately.

**Proposed Interventions**

Table 2. Interventions proposed

| Interventions | Number of publications | Number of publications that tested an intervention | Number of publications where intervention was not rejected after a test |
|---|---|---|---|
| **Specific** | | | |
| Information literacy | 9 | 3 | 2 |
| Corrective information campaigns | 5 | 4 | 3 |
| Content/account moderation | 5 | 1 | 1 |
| Labeling | 4 | 1 | 1 |
| Security/verification | 2 | | |
| **Broad or Other** | 12 | 2 | 2 |

Note: *n*=19 publications that proposed any intervention. For examples and references, see the Proposed interventions section.

The most commonly proposed interventions to address the problem of the spread and influence of manipulative information about climate change include information literacy (47%), corrective information campaigns (26%), and content/account moderation (26%) (Table 2). Out of 19 publications that proposed any interventions, corrective information campaigns and information literacy were proposed most often and not rejected following a test using empirical data. For example, Vraga et al. (44) used an experimental design on a national sample of 1,005 participants in total to test how two types of corrective information, logic-focused (describing the rhetorical flaw of oversimplification in a problematic message) and fact-focused information corrections (focusing on scientific facts), affect the perception of manipulative information by Instagram users (also see Fig 11). Fact-focused corrections countered manipulative information by providing recipients with accurate information, while logic-focused corrections commonly highlighted the rhetorical flaw underpinning manipulations. According to the authors, both types of corrections helped reduce audiences' 'misperceptions' regarding $CO_2$ emissions, partly by decreasing the credibility of original posts containing misinformation (44).

We summarize the literature advocating interventions over specific types of variables in Fig 8. It shows that corrective information has a potential for tackling manipulation exacerbated by personal traits, beliefs, or ideology. Information literacy campaigns can also influence the latter variables; such campaigns can also be used to address political and social contextual factors contributing to the spread of manipulation.

Fig 8. Distribution of variables exacerbating the diffusion of manipulative information about climate change over combinations of interventions

Notes: *n*=21; we excluded eligible studies with no identifiable variables; if one publication reported several variables or proposed several interventions, we counted them separately.

Fig 9 shows how proposed interventions depend on the format of the studied manipulation: content or account moderation seems to be the most promising intervention if manipulative information is presented in an audio-visual format. At the same time, information literacy campaigns can address manipulative content presented in a textual or visual format. For instance, Anderson and Becker (36) suggested adding humor and sarcasm factors to YouTube video content related to climate change issues to engage more audiences who have a low interest in this field. Twitter and YouTube were platforms on which the studies proposing an intervention focused, and roughly the same types of interventions were recommended for both, despite multiple differences in affordances and design (Fig 10). In another study, Buchanan et al. (45) advocated using content moderation by removing potential manipulation with the help of an algorithmic approach. It is important to highlight that the majority of the reviewed studies did not test the interventions they proposed.

Fig 9. Distribution of manipulation format over combinations of proposed interventions

Notes: *n*=19 publications that proposed any intervention; we excluded eligible studies with no identifiable variables; if one publication reported on several formats or proposed several interventions, we counted them separately.

Fig 10. Distribution of platform types over combinations of proposed interventions

Notes: *n*=19 publications that proposed any intervention; we excluded eligible studies with no identifiable variables; if one publication reported on several platforms or proposed several interventions, we counted them separately.

Half of the publications (50%) analyzed did not propose any intervention, or proposed interventions that were too broad. For example, Sanford et al. (46) suggested that scientists communicating climate change-related information 'need to deepen their understanding of how landmark, science-based reports are communicated and discussed on social media' without providing specific suggestions. Williams et al. (7) advocated for platforms to create a better environment with diverse discussions for 'mixed-attitude communities' to address a 'stabilising effect by reinforcing existing views' from echo chambers. These are important suggestions which, however, do not offer specific recommendations that platform companies or regulators can implement. Overall, to date, much of the research does not offer or evaluate specific measures for redressing digital manipulation about climate change.

Those few publications in our sample that reported the causal evidence for relevant interventions they proposed are summarized in Fig 11. We relied on a list of causal inference techniques to determine whether a publication reported causal evidence derived from a causal inference technique commonly considered in similar systematic literature review methodologies (see an example in Lorenz-Spreen et al. (29), Fig 7). We did not pursue a traditional meta-analysis design as the sub-sample of causality-reporting publications was too low. This summary highlights that interventions involving corrective content and priming of critical thinking can reduce the misperception of information about climate change across platforms such as Facebook, Twitter, and Instagram. In particular, priming critical thinking seemed to have a promising effect on reducing the trust and spread of manipulative content related to climate change (47).

Fig 11. Causal evidence summary

Notes: T: treatment; O: outcome; H: sources of effect heterogeneity or moderators; +: positive effect; -: negative effect Green: effective/beneficial interventions; Blue: measures open to interpretation.

Fig 11 also shows that the effectiveness of specific corrective strategies is contextual and depends on heterogeneous factors such as platforms, the time of placing corrections (44), audiences' pre-existing misperceptions of climate change (48) and political ideology (49). For example, logic-focused corrective content, e.g., refuting rhetorical and logical flaws in the original post, helped to reduce audiences' misperceptions 'regardless of the placement before or after the misinformation' on Instagram (Vraga et al., 2020, p. 632). At the same time, fact-focused corrective content only worked when it was placed after misinformation. In another study focusing on Twitter (48), the effects of logic-based and humor-based corrections were mediated by audiences' pre-existing attitudes toward climate change. In addition, these two intervention strategies were more effective if the issue in question was health-related rather than for climate change misinformation. Lawrence and Estow (2017) also found a potential effect of corrective comments on reducing misperceptions among people with consistent liberal beliefs on Facebook. However, they emphasized the importance of collaboration comments involving various types of stakeholders to counter misinformation, as collaboration is a promising type of intervention regardless of political orientation, according to the authors.

## Discussion

To understand the information aspect of climate denial, a great deal of work must be done to analyze how problematic content spreads on social media—a key source of information about climate change. One aspect makes the misleading content about climate change stand out as an applied context compared to other domains of misinformation: such content is often propagated by established and well-known entities with an economic interest in certain perceptions of climate change, along with political and social radical forces. The former are often interested in a rather single-stranded agenda of profiteering despite or through climate-related policies related to such domains as fossil fuel or air pollution. These entities emerge as sources of manipulation because they might confuse the debates and make them more blurred, creating uncertainty about scientific facts. Hence, we could not simply rely on 'misinformation,' 'disinformation,' or 'climate denial' as the only concepts that summarize the problem. Instead, we have extended our review of all such information and possible elements of climate change communication under the broader umbrella of manipulation.

The literature directs us towards a focus on long-term solutions, such as corrective information campaigns or information literacy promotion, to reduce the 'misperception' of climate change information. This means that the audiences of social media should be warned about lies and manipulation they might encounter; they also should be given a chance to learn how to avoid them by using online media more effectively and in a more literate way. Our study clearly highlights media literacy and corrective information as interventions that can reduce the misperception of information about climate change. However, the results are highly context-dependent. These findings echo other systematic reviews on solutions to the general problem of misinformation on social media (27,30). However, our study is the first one to emphasize the importance of these interventions in the literature concerning climate change manipulation based on a rigorous review of a relatively large body of literature.

Moreover, we found that relatively few publications suggested account moderation or suspension as an intervention—an approach with effects that have been uncertain in other studies. In addition, security or verification-linked interventions that would involve additional user verification or de-anonymization policies on a platform were only offered in two papers. There seems to be insufficient attention to this type of intervention. One area for future research could be to see which types of intervention generalizable and which ones are too context-specific to allow applying to different contexts.

We observed that many analyzed publications did not specify clearly enough some of the important elements of their research design or results, such as the exact source or format of manipulation they analyzed. Most publications also did not test the interventions they proposed, which reduced the reliability of the results. Hence, we urge authors of publications proposing solutions and interventions to mitigate the spread and impact of climate change manipulation to consider specific rather than broader recommendations. Experimental or quasi-experimental design emerging as a common way to reliably test some design interventions such as labeling, corrective information campaigns, or security/verification (examples from health communication (30) and politics (50) can provide the way forward), while longitudinal designs used in education research would be required to test information literacy campaign effectiveness (51). By addressing this design and reporting issue, studies into climate change manipulation can help us learn more about mitigating its consequences.

We also found several important research gaps, including the lack of attention to large commercial and political forces involved in manipulation generation and spreading. Indeed, we might know something about the recipients—attributes and actions of the deceived, but less about the senders—attributes and actions of deceptive entities (13), specifically those entities involved in large-scale coordinated campaigns of climate information distortions. Moreover, the reviewed publications paid relatively little attention to video- and image-focused platforms whose audiences have been growing rapidly in recent years. Finally, we found that researchers used multiple research methods, such as surveys, interviews, and data science, to answer some of these important questions. Nevertheless, the potential of methods that rely on large quantities of data seems to be underused in the body of literature.

Finally, this review has focused on manipulative information. Perhaps the research we have reviewed can also be used to tackle the other side of the coin: promoting truthful or reliable information. That might include placing correct information next to manipulative information, perhaps in automated ways, or fostering online groups (or 'communities') that promote sharing or producing reliable information, including information that speaks to those affected by climate change where such information is likely to be most effective. Social media that are not gatekept and their ill effects have been a large focus of recent research. However, they and other sources of online information (like Wikipedia) can equally have advantages over traditional media or offline information in promoting more useful and reliable knowledge.

## Conclusion

We have found that research into the spread of manipulative information about climate change on social media is rather fragmented, with many country-based or platform-based case studies. Taking a bird's eye view of this research allows us to note a continued skew towards social media that are less widely used for climate change information—or manipulation—dissemination than more prominent ones. We can also see that the interventions or policy implications remain rather general. Furthermore, there is a focus on the US and its recent politics, lessons from which may be hard to carry over into other contexts. If we compare the social media research in our study with those that are actually used (5), then there should be a major realignment in the research. However, even that would not be an optimal realignment since, first, the eight countries in Ejaz et al. (5) are not representative of global media uses, and second, there is a huge skew towards research on text, whereas images and video are used far more widely. This points out how a theoretical synthesis is required for the more complex environment that has emerged both politically and in terms of social media usage.

In the future, automated methods are bound to become more central to this research. Our research informs these methods by directing them toward the most high-impact interventions. Perhaps more importantly, it can be envisioned that a systematic link could be attempted between findings and interventions, since automated methods allow for examining the feedback loops between online climate information and manipulation and their effects. In this respect, social scientists are now competing with the private sector and its marketing techniques, which are able to use tools to shape public understanding of climate change, and where the border between marketing and manipulation is becoming blurred.

This study is the first systematic review specifically focusing on the spread of manipulative information about climate change on social media. Other studies can expand our review once more evidence emerges about the influence, spread, and social contexts of this problem. It is possible that some studies were excluded from the systematic review even though we undertook an extensive search that included two large social science databases and numerous search terms. Still, with our relatively broad search strategy, we had to find a boundary where we would stop our selection. Nevertheless, we systematically proceeded with the studies that do provide empirical evidence about how to respond to online manipulations with effective mitigating strategies. There are relatively few studies that expanded on the subject of exaggeration of manipulation spread or tested proposed interventions. With the growing body of literature on this important topic, we hope to be able to update our research based on newer evidence. In the meantime, this study should provide an important guide to and outlook for the state of the rapidly expanding field of online manipulation about climate change.

**Figures**

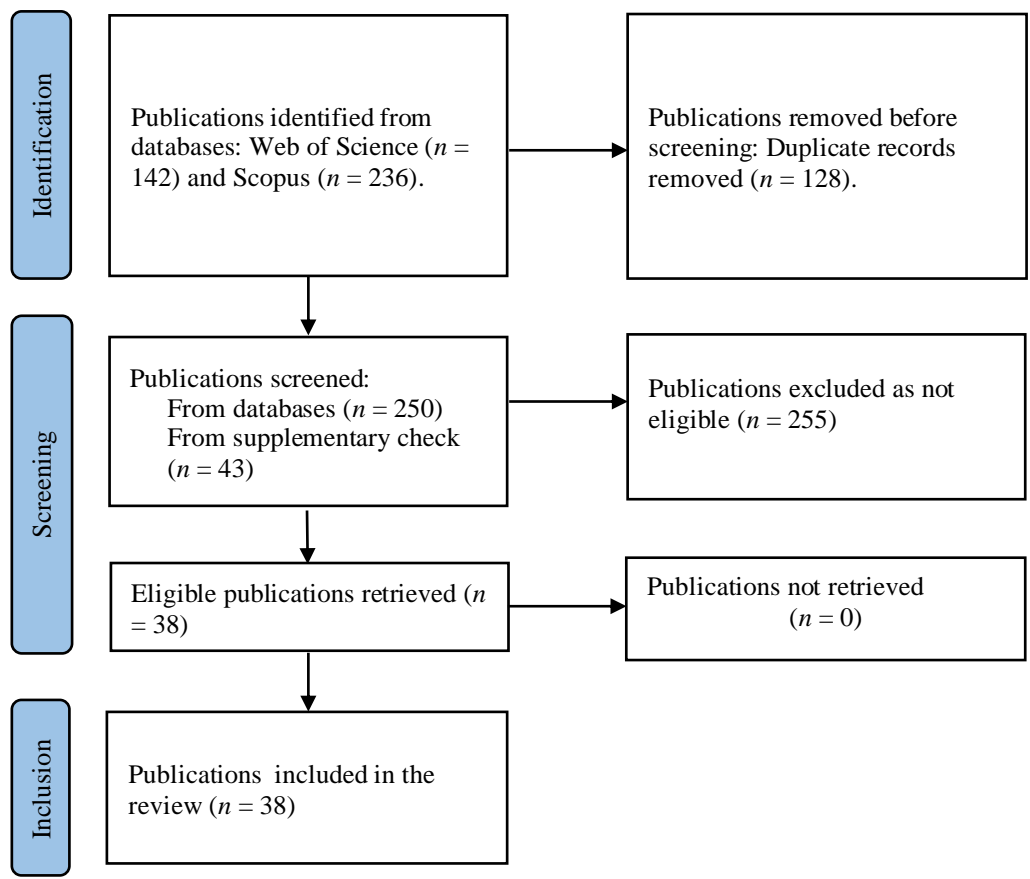

Fig 1. Flow of publications through different stages of the systematic review

Fig 2. Temporal trends in eligible publications

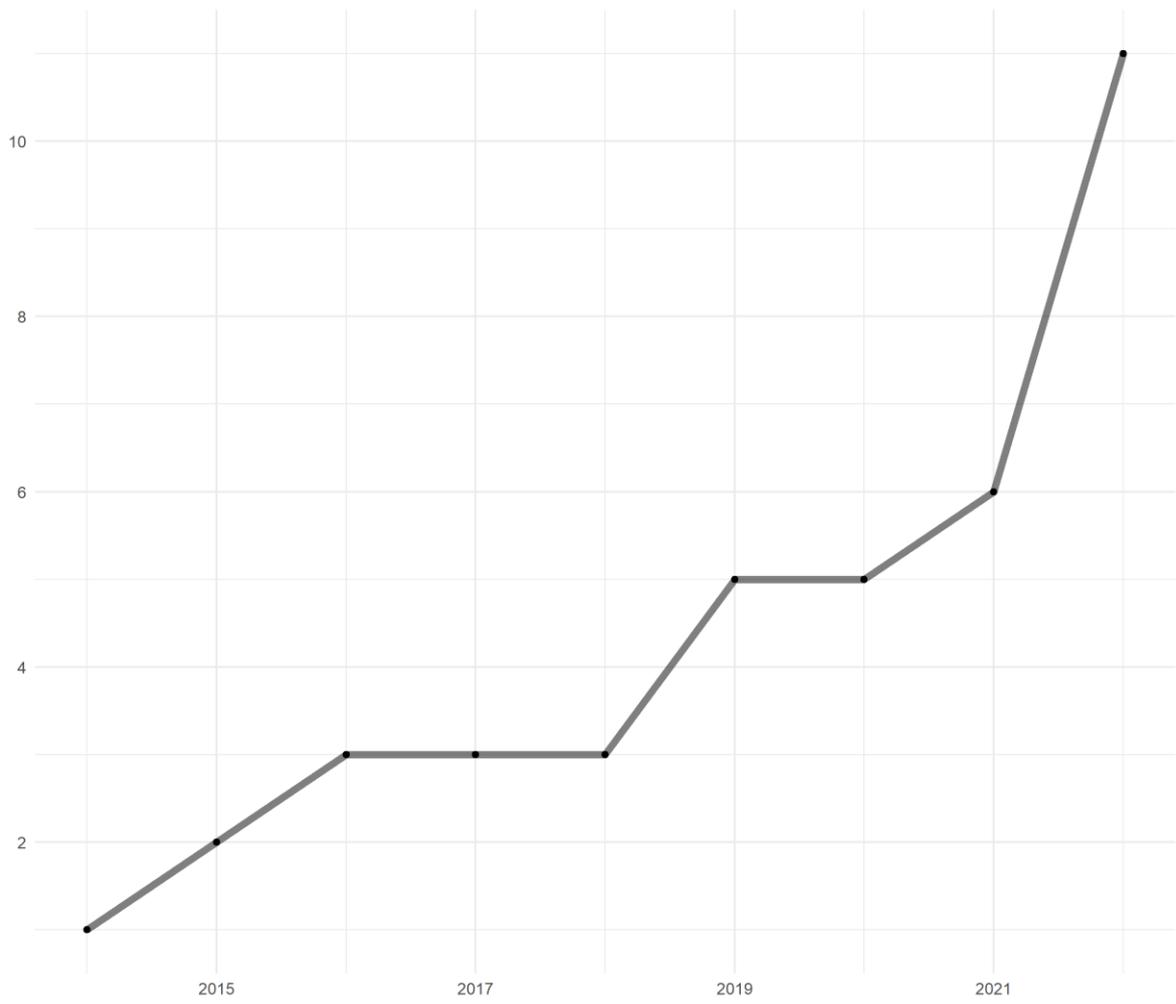

Note: *n*=38 publications.

Fig 3. Country of affiliation for the first authors of the eligible publications

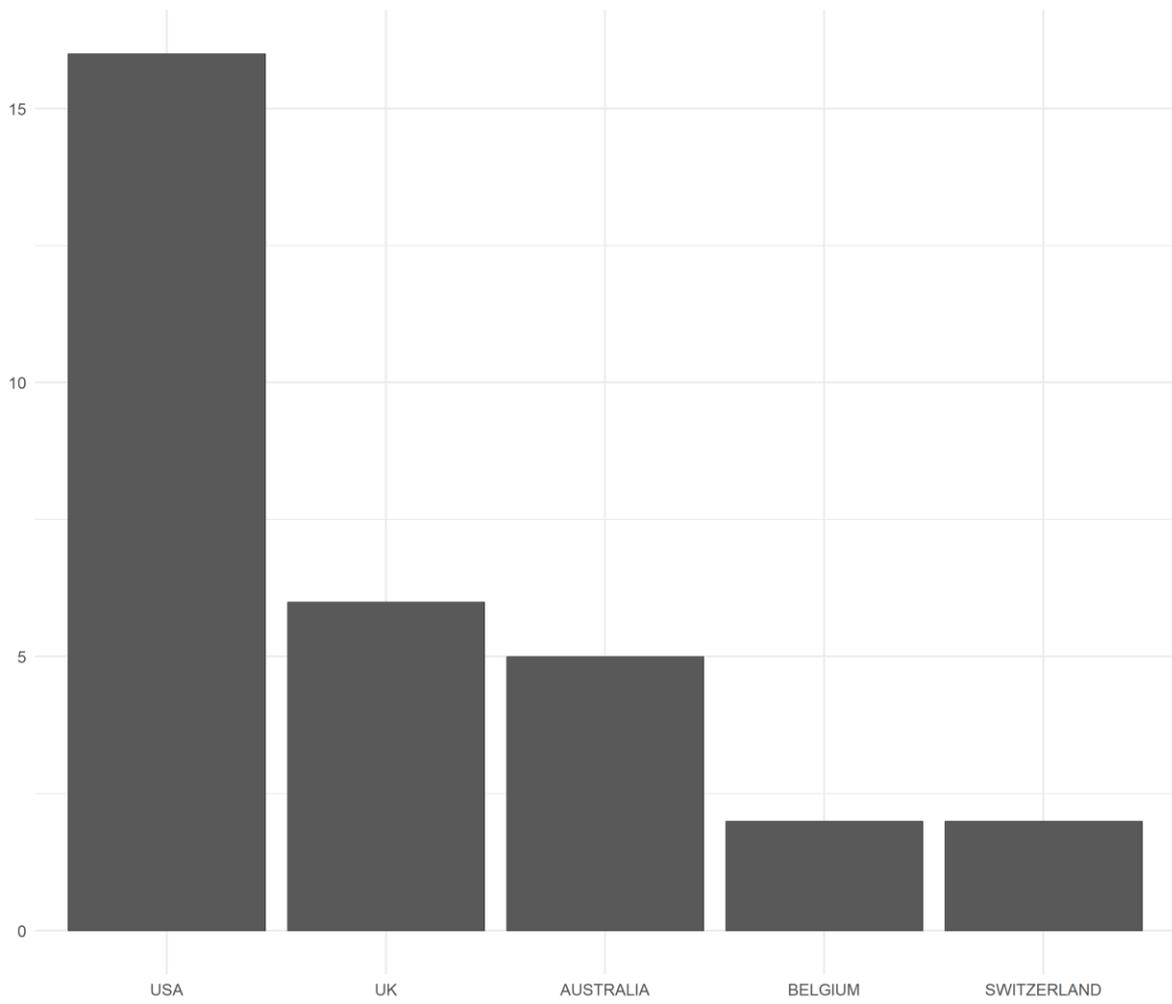

Note: *n*=38 publications.

Fig 4. Methods used by the analyzed publications

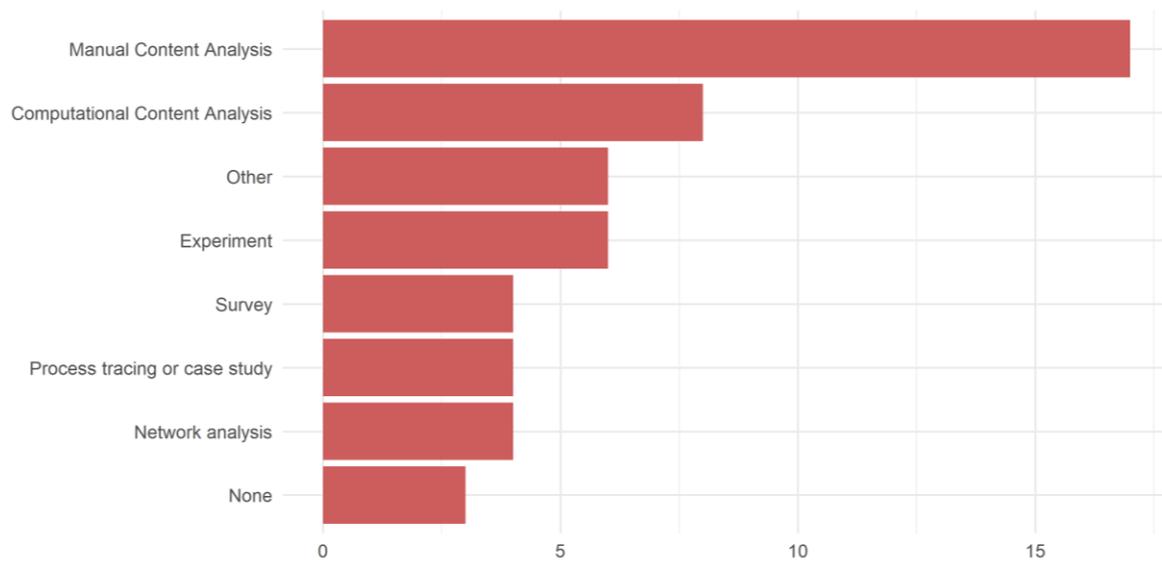

Note: *n*=38 publications.

Fig 5. Sources of manipulation studied

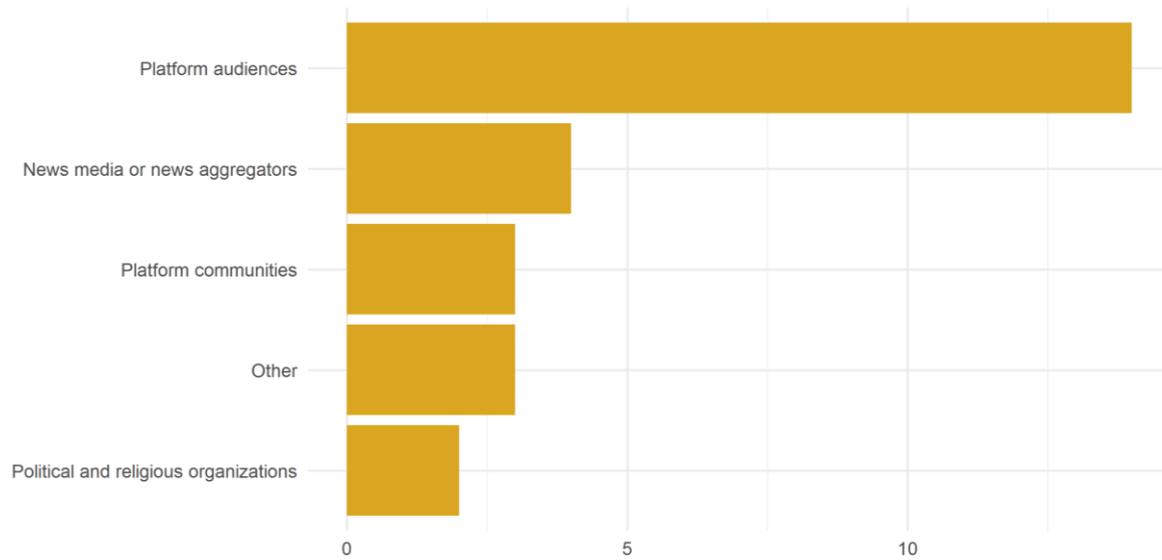

Note: *n*=26 publications that identified or proposed the source of manipulation. Communities refer to defined or organized units on platform, such as groups or pages, while audiences refer to broader and undefined cohorts of users.

Fig 6. Variables exacerbating the diffusion of manipulative information about climate change

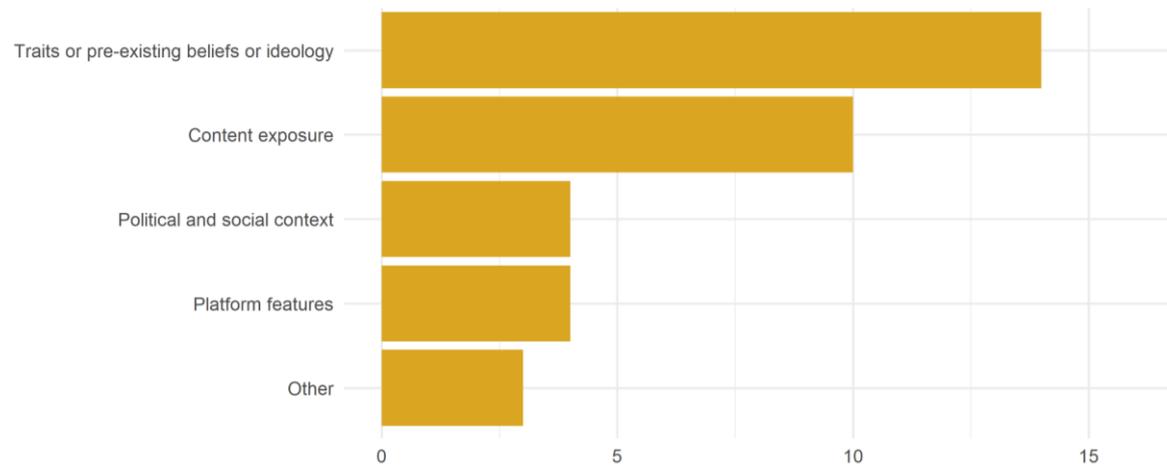

Note: *n*=21; we excluded eligible studies with no identifiable variables.

Fig 7. Distribution of variables exacerbating the diffusion of manipulative information about climate change over combinations of methods

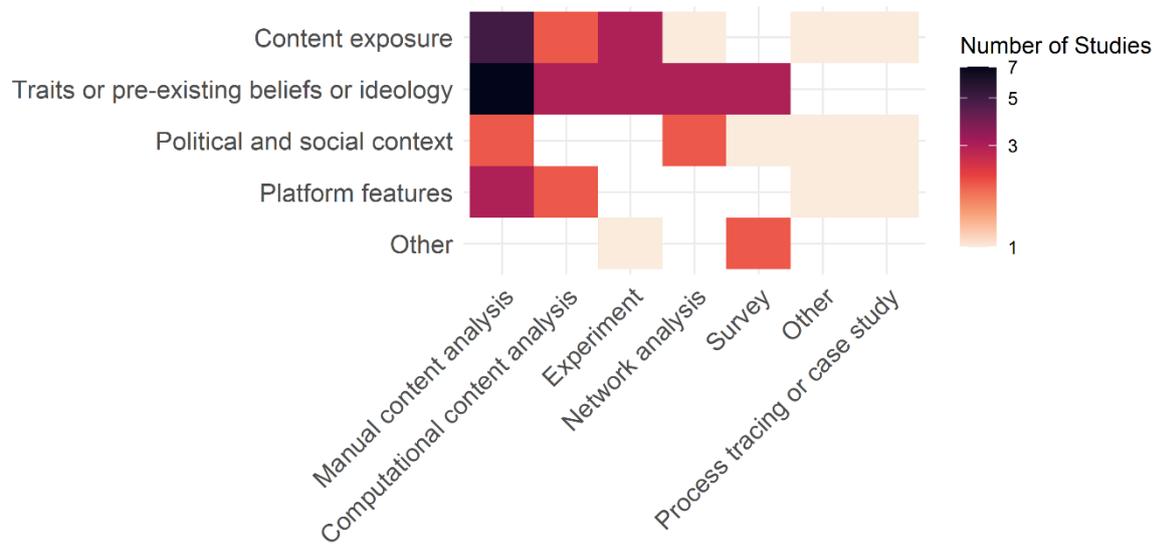

Notes: *n*=21; we excluded eligible studies with no identifiable variables; if one publication reported several variables or used several methods, we counted them separately.

Fig 8. Distribution of variables exacerbating the diffusion of manipulative information about climate change over combinations of interventions

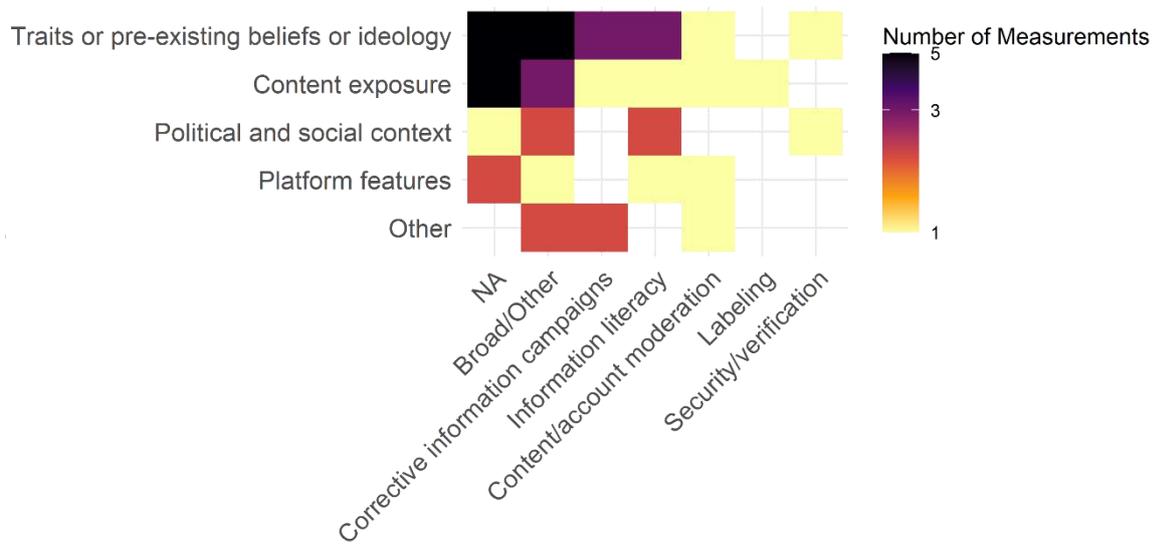

Notes: *n*=21; we excluded eligible studies with no identifiable variables; if one publication reported several variables or proposed several interventions, we counted them separately.

Fig 9. Distribution of manipulation format over combinations of proposed interventions

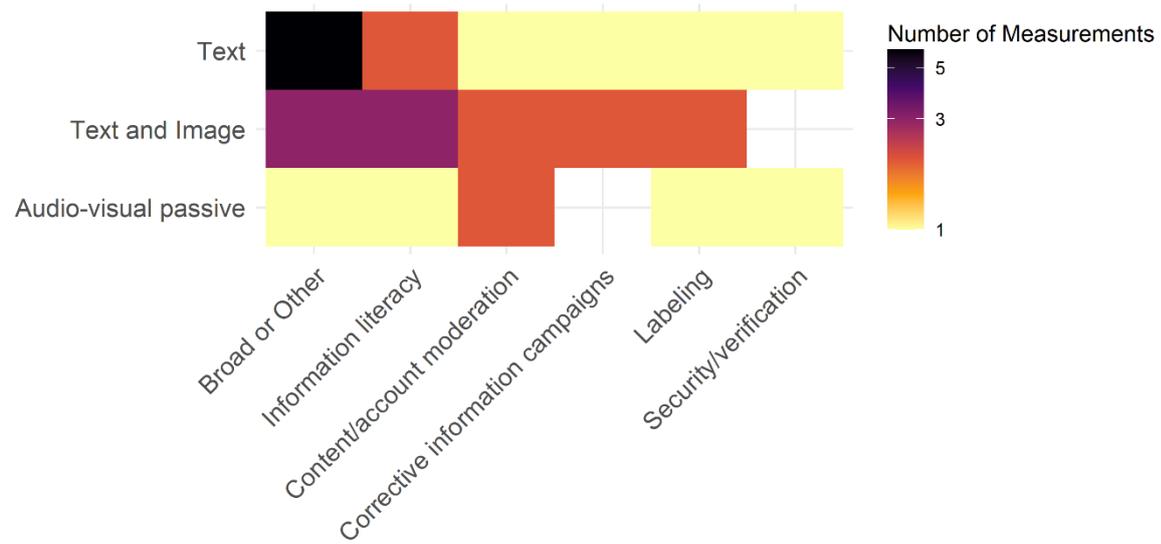

Notes: *n*=19 publications that proposed any intervention; we excluded eligible studies with no identifiable variables; if one publication reported on several formats or proposed several interventions, we counted them separately.

Fig 10. Distribution of platform types over combinations of proposed interventions

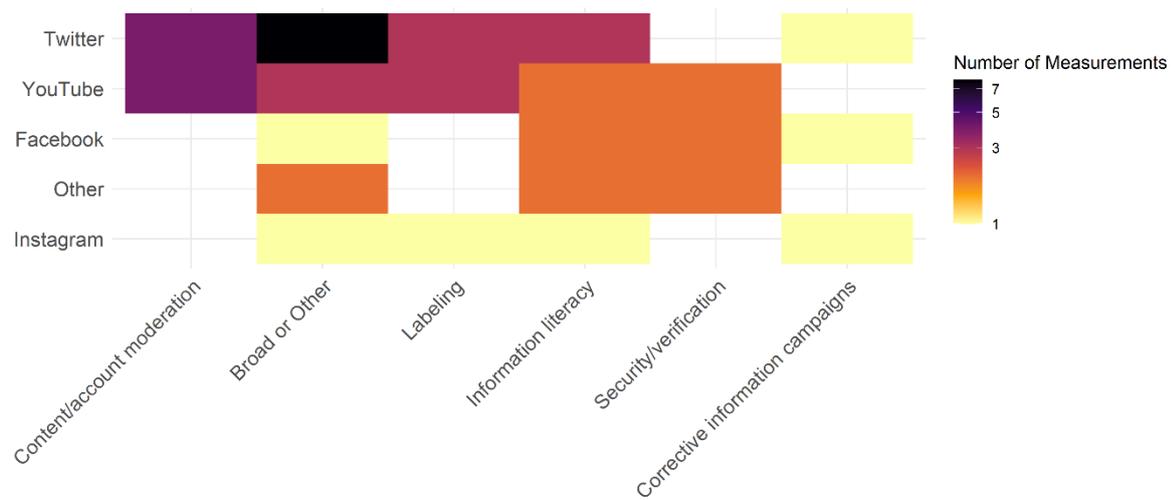

Notes: *n*=19 publications that proposed any intervention; we excluded eligible studies with no identifiable variables; if one publication reported on several platforms or proposed several interventions, we counted them separately.

Fig 11. Causal evidence summary

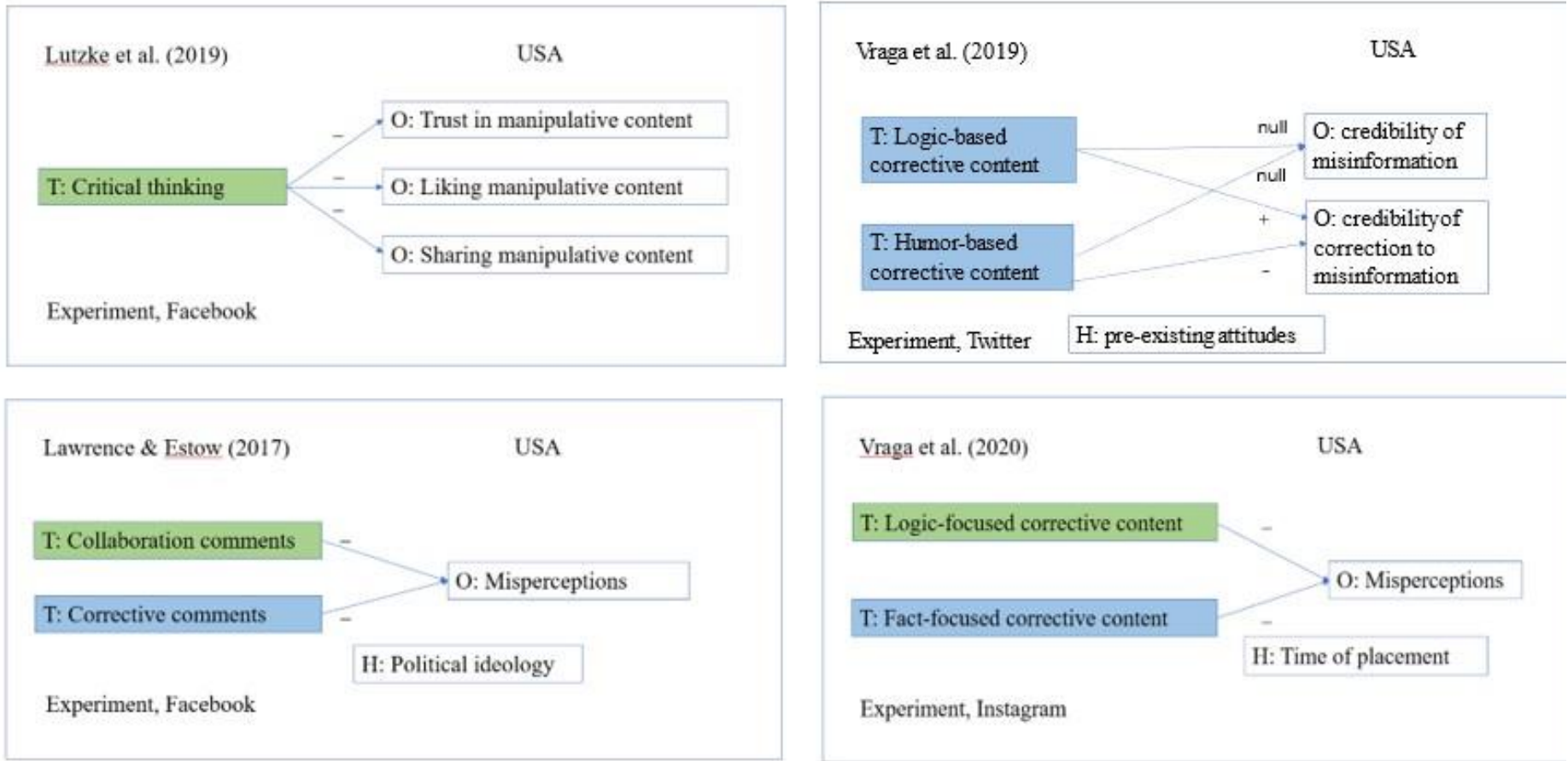

Notes: T: treatment; O: outcome; H: sources of effect heterogeneity or moderators; +: positive effect; -: negative effect Green: effective/beneficial interventions; Blue: measures open to interpretation.

# Supporting Information

**Supplementary Table S1. Codebook.**

| Code | Description of the code | Variables |
|---|---|---|
| Eligibility criteria | | |
| English | Publication is available in English. | 0 = no<br>1 = yes |
| Manipulation | Study discusses any aspects of misinformation, disinformation, fake news, propaganda, 'credibility' of information or digital/automated manipulation as a key object of study. If there are several objects, manipulation should be one of at least three of them. | 0 = no<br>1 = yes |
| Digital platforms | Social media platforms (websites that are dependent primarily on user-generated content that facilitates two-way interaction) constitute a significant focus of the research. The study names any social media platform as a major study object or its key background, field, or context. Forums are not considered social media. | 0 = no<br>1 = yes |
| Climate change domain | A study explores manipulation in relation to climate change. By the climate change domain, we mean any study that explores human-induced global warming and human influence on the global climate. | 0 = unrelated<br>1 = climate change domain |

Full paper coding criteria [coded only if all the Eligibility variables are '1']

| | | |
|---|---|---|
| Methods | Methods used to collect data (multiple selection coding category). | 0 = None<br>1 = Survey (real world)<br>2 = Interview<br>3 = Focus group<br>4 = Ethnography/Observation<br>5 = Experiment<br>6 = Content/textual analysis (manual)<br>7 = Content analysis (automated)/Social media data extraction<br>8 = Network analysis<br>9 = Agent-based modeling & simulation<br>10 = Process tracing and/or case study<br>11 = Review studies (like syst review)<br>12 = non-empirical essays, comments<br>50 = Other |
| Platform | What social media platform does this study focus on? Can include several options. If it mentioned multiple platforms, select only the main ones for the study. | 0 = Unclear/No specific platform<br>1 = Facebook<br>2 = Twitter<br>3 = Telegram<br>4 = YouTube<br>5 = Instagram<br>6 = WhatsApp |
| Manipulation source | Specify manipulation source(s) in relation to climate change discussed in the study. | 7 = Other<br>[open coding] |
| Manipulation format | How was the studied manipulative information/campaign formatted? | 0 = No particular media context/unclear<br>1 = Text<br>2 = Text and Image<br>3 = Audio-visual passive<br>4 = Other format(s) |
| Channel | How does the studied manipulation spread? Specify the channel(s) used to spread manipulation in relation to climate change discussed in the study. | [open coding] |
| Variables | If the publication studied/found variables that exacerbate the diffusion of climate change manipulative content, describe them. | [open coding] |
| Findings | Specify the findings of the paper in a few words in relation to the effects of the studied manipulation of climate change. | [open coding] |
| Narratives | Specify key manipulation narratives that are discussed in the study in relation to climate. | [open coding] |
| Causal evidence report | Does the study report causal evidence derived from causal inference? Use the summary of the main techniques of casual inference as presented in Figure 7 in Lorenz-Spreen et al. (2022). | 0 = no<br>1 = yes |
| | Other relevant techniques, e.g. field experiments, are also considered. | |
| Interventions | Interventions mitigating the impact of misinformation proposed in the publication. Use examples to determine. If nothing is proposed, put NA and do NOT proceed with the variables of the section proposed countermeasures. Examples:<br>0 broad: unclear/not specific/not platform related. Example: 'culture of humility', 'misinformation detection' (not clear how and why). If two broad, use the next column to specify:<br>1 Advertisement policy: If the intervention changes the advertisement policy of the platform and has a user-facing component, e.g., (i) Facebook requires the 'Paid for by' label, (ii) Facebook has an information button for advertisements.<br>2 Content labeling: If the intervention labels posts, accounts, stories with (i) a fact-checking tag, (ii) funding/advertising tag, (iii) outdated tag, or any other forms of tagging, including providing further context without the user having to click through to receive the additional information; e.g., (i) Facebook adds fact-check labels to posts, (ii) Twitter labels tweets by state-media, etc.<br>3 Content/account moderation: If the intervention involves one of the following actions: (i) Takedown: removes content/posts/accounts (takedowns), (ii) Suspension: suspends accounts/blocks accounts, (iii) AI: modifies feed, trends, content appearances, and order (algorithmic and AI changes included); e.g., (i) YouTube downranks unauthoritative content, (ii) Twitter reduces interactions with accounts that users don't follow, etc. | NA = nothing proposed<br>0 = Broad<br>1 = Advertisement policy<br>2 = Labeling<br>3 = Content/account moderation<br>4 = Content reporting<br>5 = Content distribution/sharing<br>6 = Corrective information campaigns/messages<br>7 = Disinformation disclosure<br>8 = Information literacy<br>9 = Redirection<br>10 = Security/verification<br>11 = Other |

4 Content user reporting: If the intervention changes how users report problematic content on the platform. E.g., (i) TikTok introduced a 'misinformation' option in the content reporting options.

5 Content user sharing: If the intervention targets the distribution of problematic content on platforms by users. E.g., (i) if WhatsApp limits forwards, (ii) if Pinterest prevents pinning or saving posts.

6 Corrective information campaigns/messages: example: having governments or companies publicly debunk a rumor about them on social media in a separate, unlinked piece of content; users can also be involved

7 Disinformation disclosure: If the intervention informs a user they have come in contact, shared, or interacted with disinformation; e.g., (i) Reddit telling users they've interacted with misinformation.

8 Information/media literacy: If the intervention aims to educate users to identify disinformation (or misinformation) online; e.g., (i) Snapchat's myth-busting game, (ii) Facebook's tool to help users identify misinformation.

9 Redirection/debunking: If the intervention redirects users to different information, accounts, posts, either by taking them to a different link or by offering in-app notices or if the intervention imparts and curates accurate information (including but not limited to COVID-19); e.g., (i) Instagram showing content from CDC and WHO when users search for COVID-19, (ii) Facebook and Twitter's U.S. election or COVID information hubs.

| | | |
|---|---|---|
| | 10 Security/verification: If the intervention increases or decreases the security or verification requirements on the platform; e.g., (i) Twitter's protection program for political officials.<br>11 Other. Something related to platform work but not mentioned here. | |
| Interventions Specify | Shortly describe the proposed interventions that are related to the functioning of platforms. | [open coding] |
| Interventions Empirical | Does the publication demonstrate any attempt to test the impact/effectiveness/influence of intervention(s) and preventive strategies it examines empirically? The publication should state whether and how these interventions had been tested, not necessarily in a casual manner, empirical evidence is required. | 0 = no<br>1 = yes |
| Interventions Empirical Outcome | If an intervention was empirically tested, specify if the intervention confirmed or rejected the positive impact of the intervention on manipulation reduction. Multichoice is possible. | 0 = rejected<br>1 = confirmed<br>2 = uncertain |
| Interventions Empirical Specify | Provide details on which measures were confirmed and rejected as a result of empirical tests based on the examples for the Inters variable. | [open coding] |
| Notes | If uncertain or found something unusual, please note here. | [open coding] |

**Supplementary Table S2. Publications that met the eligibility criteria**

| Authors | Publication Title | DOI or URL |
|---|---|---|
| AL-RAWI A;OKEEFE D;KANE O;BIZIMANA A | TWITTERS FAKE NEWS DISCOURSES AROUND CLIMATE CHANGE AND GLOBAL WARMING | https://doi.org/10.3389/fcomm.2021.729818 |
| ANDERSON A;BECKER A | NOT JUST FUNNY AFTER ALL: SARCASM AS A CATALYST FOR PUBLIC ENGAGEMENT WITH CLIMATE CHANGE | https://journals.sagepub.com/doi/10.1177/1075547018786560 |
| ANDERSON A;HUNTINGTON H | SOCIAL MEDIA SCIENCE AND ATTACK DISCOURSE HOW TWITTER DISCUSSIONS OF CLIMATE CHANGE USE SARCASM AND INCIVILITY | https://doi.org/10.1177/1075547017735113 |
| ARLT D;HOPPE I;SCHMITT J;DE S F;BRÜGGEMANN M | CLIMATE ENGAGEMENT IN A DIGITAL AGE EXPLORING THE DRIVERS OF PARTICIPATION IN CLIMATE DISCOURSE ONLINE IN THE CONTEXT OF COP21 | https://doi.org/10.1080/17524032.2017.1394892 |
| BEDNAREK M;ROSS A;BOICHAK O;DORAN Y;CARR G;ALTMANN E;ALEXANDER T | WINNING THE DISCURSIVE STRUGGLE THE IMPACT OF A SIGNIFICANT ENVIRONMENTAL CRISIS EVENT ON DOMINANT CLIMATE DISCOURSES ON TWITTER | https://doi.org/10.1016/j.dcm.2021.100564 |
| BESSI A;ZOLLO F;DEL V M;SCALA A;CALDARELLI G;QUATTROCIOCCHI W | TREND OF NARRATIVES IN THE AGE OF MISINFORMATION | https://doi.org/10.1371/journal.pone.0134641 |

| Authors | Title | Link |
|---|---|---|
| BHUIYAN M;ZHANG A;SEHAT C;MITRA T | INVESTIGATING DIFFERENCES IN CROWDSOURCED NEWS CREDIBILITY ASSESSMENT RATERS TASKS AND EXPERT CRITERIA | https://dl.acm.org/doi/abs/10.1145/3415164 |
| BLOOMFIELD E; TILLERY D | THE CIRCULATION OF CLIMATE CHANGE DENIAL ONLINE: RHETORICAL AND NETWORKING STRATEGIES ON FACEBOOK | https://www.tandfonline.com/doi/full/10.1080/17524032.2018.1527378 |
| BUCHANAN G;KELLY R;MAKRI S;MCKAY D | READING BETWEEN THE LIES A CLASSIFICATION SCHEME OF TYPES OF REPLY TO MISINFORMATION IN PUBLIC DISCUSSION THREADS | https://doi.org/10.1145/3498366.3505823 |
| CALYX C;LOW J | HOW A CLIMATE CHANGE SCEPTIC POLITICIAN CHANGED THEIR MIND | https://doi.org/10.22323/2.19030304 |
| CANN T; WEAVER I; WILLIAMS H | IDEOLOGICAL BIASES IN SOCIAL SHARING OF ONLINE INFORMATION ABOUT CLIMATE CHANGE | https://journals.plos.org/plosone/article?id=10.1371/journal.pone.0250656 |
| CHEEMA G;HAKIMOV S;SITTAR A;MULLER-BUDACK E;OTTO C;EWERTH R | A DATASET FOR MULTIMODAL CLAIM DETECTION IN SOCIAL MEDIA | https://aclanthology.org/2022.findings-naacl.72/ |
| CHEN C; SHI W; YANG J; FU H | SOCIAL BOTS' ROLE IN CLIMATE CHANGE DISCUSSION ON TWITTER: MEASURING STANDPOINTS, TOPICS, AND INTERACTION STRATEGIES | https://www.sciencedirect.com/science/article/pii/S1674927821001490?via%3Dihub |
| COOK J;BEDFORD D;MANDIA S | RAISING CLIMATE LITERACY THROUGH ADDRESSING MISINFORMATION CASE STUDIES IN AGNOTOLOGYBASED LEARNING | https://doi.org/10.5408/13-071.1 |

| Authors | Title | Link |
|---|---|---|
| DIEHL T; HUBER B; DE-ZUNIGA H; LIU J | SOCIAL MEDIA AND BELIEFS ABOUT CLIMATE CHANGE: A CROSS-NATIONAL ANALYSIS OF NEWS USE, POLITICAL IDEOLOGY, AND TRUST IN SCIENCE | https://academic.oup.com/ijpor/article/33/2/197/5628296?login=true |
| ESLEN-ZIYA H | HUMOUR AND SARCASM: EXPRESSIONS OF GLOBAL WARMING ON TWITTER | https://www.nature.com/articles/s41599-022-01236-y |
| FINE J;LOVE-NICHOLS J | WE ARE NOT THE VIRUS COMPETING ONLINE DISCOURSES OF HUMANENVIRONMENT INTERACTION IN THE ERA OF COVID19 | https://doi.org/10.1080/17524032.2021.1982744 |
| FORTI L;TRAVASSOS M;CORONEL-BEJARANO D;MIRANDA D;SOUZA D;SABINO J;SZABO J | POSTS SUPPORTING ANTIENVIRONMENTAL POLICY IN BRAZIL ARE SHARED MORE ON SOCIAL MEDIA | https://doi.org/10.1007/s00267-022-01757-x |
| GILLAM C | FINDING AND FOLLOWING THE FACTS IN AN ERA OF FAKE NEWS | https://doi.org/10.4324/9781351068406-7 |
| JACQUES P;KNOX C | HURRICANES AND HEGEMONY A QUALITATIVE ANALYSIS OF MICROLEVEL CLIMATE CHANGE DENIAL DISCOURSES | https://doi.org/10.1080/09644016.2016.1189233 |
| LAWRENCE E;ESTOW S | RESPONDING TO MISINFORMATION ABOUT CLIMATE CHANGE | https://doi.org/10.1080/1533015X.2017.1305920 |
| LUTZKE L;DRUMMOND C;SLOVIC P;ÃRVAI J | PRIMING CRITICAL THINKING SIMPLE INTERVENTIONS LIMIT THE INFLUENCE OF FAKE NEWS ABOUT CLIMATE CHANGE ON FACEBOOK | https://doi.org/10.1016/j.gloenvcha.2019.101964 |

| Authors | Title | DOI/URL |
|---|---|---|
| LYDIRI M;EL H Y;ZOUGAGH H | SENTIMENT ANALYSIS DECISION SYSTEM FOR TRACKING CLIMATE CHANGE OPINION IN TWITTER | https://doi.org/10.1007/978-3-031-06458-6_15 |
| MAZID M;ZARNAZ Z | CLIMATE CHANGE MYTHS DETECTION USING DYNAMICALLY WEIGHTED ENSEMBLE BASED STANCE CLASSIFIER | https://doi.org/10.1145/3542954.3542995 |
| MOERNAUTA R; MAST J; TEMMERMAN M; BROERSMA M | HOT WEATHER, HOT TOPIC. POLARIZATION AND SCEPTICAL FRAMING IN THE CLIMATE DEBATE ON TWITTER | https://www.tandfonline.com/doi/full/10.1080/1369118X.2020.1834600 |
| MOROSOLI S;VAN A P;HUMPRECHT E;STAENDER A;ESSER F | IDENTIFYING THE DRIVERS BEHIND THE DISSEMINATION OF ONLINE MISINFORMATION A STUDY ON POLITICAL ATTITUDES AND INDIVIDUAL CHARACTERISTICS IN THE CONTEXT OF ENGAGING WITH MISINFORMATION ON SOCIAL MEDIA | https://doi.org/10.1177/00027642221118300 |
| NUCCI A;HIBBERD M | ECOACTIVISM HUMANCOMPUTER INTERACTION AND FAST FASHION | https://doi.org/10.1007/978-3-030-78224-5_36 |
| REED M | 'THIS LOOPY IDEA' AN ANALYSIS OF UKIP'S SOCIAL MEDIA DISCOURSE IN RELATION TO RURALITY AND CLIMATE CHANGE | https://www.tandfonline.com/doi/full/10.1080/13562576.2016.1192332 |
| ROSS A; RIVERS D | INTERNET MEMES, MEDIA FRAMES, AND THE CONFLICTING LOGICS OF CLIMATE CHANGE DISCOURSE | https://www.tandfonline.com/doi/full/10.1080/17524032.2018.1560347 |
| SADIK S;BENEDETTI J;GOKHALE S | ANALYZING CLIMATE CHANGE DIALOGUE DURING CALIFORNIA WILDFIRES | https://doi.org/10.1109/ASIANCON55314.2022.9908642 |

| Authors | Title | Link |
|---|---|---|
| SANFORD M; PAINTER J; YASSERI T; LORIMER J | CONTROVERSY AROUND CLIMATE CHANGE REPORTS: A CASE STUDY OF TWITTER RESPONSES TO THE 2019 IPCC REPORT ON LAND | https://link.springer.com/article/10.1007/s10584-021-03182-1 |
| TILLERY D;BLOOMFIELD E | HYPERRATIONALITY AND RHETORICAL CONSTELLATIONS IN DIGITAL CLIMATE CHANGE DENIAL A MULTIMETHODOLOGICAL ANALYSIS OF THE DISCOURSE OF WATTS UP WITH THAT | https://doi.org/10.1080/10572252.2021.2019317 |
| TINGLEY D;WAGNER G | SOLAR GEOENGINEERING AND THE CHEMTRAILS CONSPIRACY ON SOCIAL MEDIA | https://doi.org/10.1057/s41599-017-0014-3 |
| TREEN K;WILLIAMS H;O'NEILL S | ONLINE MISINFORMATION ABOUT CLIMATE CHANGE | https://doi.org/10.1002/wcc.665 |
| VRAGA E;KIM S;COOK J | TESTING LOGICBASED AND HUMORBASED CORRECTIONS FOR SCIENCE HEALTH AND POLITICAL MISINFORMATION ON SOCIAL MEDIA | https://doi.org/10.1080/08838151.2019.1653102 |
| VRAGA E;KIM S;COOK J;BODE L | TESTING THE EFFECTIVENESS OF CORRECTION PLACEMENT AND TYPE ON INSTAGRAM | https://doi.org/10.1177/1940161220919082 |
| WILLIAMS H; MCMURRAY J; KURZ T; LAMBERT F | NETWORK ANALYSIS REVEALS OPEN FORUMS AND ECHO CHAMBERS IN SOCIAL MEDIA DISCUSSIONS OF CLIMATE CHANGE | https://doi.org/10.1016/j.gloenvcha.2015.03.006 |
| XU Z;ATKIN D | FRAMING CLIMATE CHANGE IN THE 5TH ESTATE COMPARING ONLINE ADVOCACY AND DENIAL WEBPAGES AND THEIR ENGAGEMENT | https://doi.org/10.1177/19312431221087247 |

**Supplementary Table S3. Krippendorff's α and agreement**

| Category | Alpha |
|---|---|
| English | 1 |
| Manipulation | 0.78 |
| Platforms | 0.68 |
| Climate change* | 0.64 |

Note: *n*=146 publications. Krippendorff's α is sensitive to skewed rates. Domain and platform variables in our dataset were skewed as most publications were not concerned with climate or platforms. Therefore, we accepted lower rates for Krippendorff's α as sufficient.
*Variable was highly homogeneous, with most coded units falling in one category. This is a common issue with some types of coding (Lacy et al., 2015).

**Supplementary Table S4. PRISMA 2020 Checklist**

| Section and Topic | Item # | Checklist item | Location where item is reported |
|---|---|---|---|
| **TITLE** | | | |
| Title | 1 | Identify the report as a systematic review. | Title Section |
| **ABSTRACT** | | | |
| Abstract | 2 | See the PRISMA 2020 for Abstracts checklist. | Abstract Section |
| **INTRODUCTION** | | | |
| Rationale | 3 | Describe the rationale for the review in the context of existing knowledge. | Introduction Section; Previous Reviews and Research Questions Section |
| Objectives | 4 | Provide an explicit statement of the objective(s) or question(s) the review addresses. | Introduction Section |
| **METHODS** | | | |
| Eligibility criteria | 5 | Specify the inclusion and exclusion criteria for the review and how studies were grouped for the syntheses. | Throughout Methods Section; Supplementary Note SN2 |
| Information sources | 6 | Specify all databases, registers, websites, organisations, reference lists and other sources searched or consulted to identify studies. Specify the date when each source was last searched or consulted. | The first paragraph of Methods Section |
| Search strategy | 7 | Present the full search strategies for all databases, registers and websites, including any filters and limits used. | The first, second and third paragraphs of Methods Section; Supplementary Note SN1 |

| Section and Topic | Item # | Checklist item | Location where item is reported |
|---|---|---|---|
| Selection process | 8 | Specify the methods used to decide whether a study met the inclusion criteria of the review, including how many reviewers screened each record and each report retrieved, whether they worked independently, and if applicable, details of automation tools used in the process. | The fourth, fifths, and the sixth paragraphs of Methods Section |
| Data collection process | 9 | Specify the methods used to collect data from reports, including how many reviewers collected data from each report, whether they worked independently, any processes for obtaining or confirming data from study investigators, and if applicable, details of automation tools used in the process. | The fourth, fifths, and the sixth paragraphs of Methods Section |
| Data items | 10a | List and define all outcomes for which data were sought. Specify whether all results that were compatible with each outcome domain in each study were sought (e.g. for all measures, time points, analyses), and if not, the methods used to decide which results to collect. | The fourth, fifths, and the sixth paragraphs of Methods Section; Supplementary Table S1 |
|  | 10b | List and define all other variables for which data were sought (e.g. participant and intervention characteristics, funding sources). Describe any assumptions made about any missing or unclear information. | Supplementary Table S1 |
| Study risk of bias assessment | 11 | Specify the methods used to assess risk of bias in the included studies, including details of the tool(s) used, how many reviewers assessed each study and whether they worked independently, and if applicable, details of automation tools used in the process. | No quantitative synthesis conducted |
| Effect measures | 12 | Specify for each outcome the effect measure(s) (e.g. risk ratio, mean difference) used in the synthesis or presentation of results. | Supplementary Table S1 |
| Synthesis methods | 13a | Describe the processes used to decide which studies were eligible for each synthesis (e.g. tabulating the study intervention characteristics and comparing against the planned groups for each synthesis (item #5)). | The fourth, fifths, and the sixth paragraphs of Methods Section; Supplementary Table S1 |
|  | 13b | Describe any methods required to prepare the data for presentation or synthesis, such as handling of missing summary statistics, or data conversions. | The fourth, fifths, and the sixth paragraphs of |

| Section and Topic | Item # | Checklist item | Location where item is reported |
|---|---|---|---|
| | | | Methods Section; Supplementary Table S1 |
| | 13c | Describe any methods used to tabulate or visually display results of individual studies and syntheses. | The fourth, fifths, and sixth paragraphs of the Methods Section; Supplementary Table S1 |
| | 13d | Describe any methods used to synthesize results and provide a rationale for the choice(s). If meta-analysis was performed, describe the model(s), method(s) to identify the presence and extent of statistical heterogeneity, and software package(s) used. | No quantitative synthesis conducted |
| | 13e | Describe any methods used to explore possible causes of heterogeneity among study results (e.g. subgroup analysis, meta-regression). | No quantitative synthesis conducted |
| | 13f | Describe any sensitivity analyses conducted to assess robustness of the synthesized results. | No quantitative synthesis conducted |
| Reporting bias assessment | 14 | Describe any methods used to assess risk of bias due to missing results in a synthesis (arising from reporting biases). | No quantitative synthesis conducted |
| Certainty assessment | 15 | Describe any methods used to assess certainty (or confidence) in the body of evidence for an outcome. | No quantitative synthesis conducted |
| **RESULTS** | | | |
| Study selection | 16a | Describe the results of the search and selection process, from the number of records identified in the search to the number of studies included in the review, ideally using a flow diagram. | Fig 1 |
| | 16b | Cite studies that might appear to meet the inclusion criteria, but which were excluded, and explain why they were excluded. | The fourth, fifths, and the sixth paragraphs of the Methods Section; |

| Section and Topic | Item # | Checklist item | Location where item is reported |
|---|---|---|---|
| | | | Supplementary Table S1 |
| Study characteristics | 17 | Cite each included study and present its characteristics. | Supplementary Table S2 |
| Risk of bias in studies | 18 | Present assessments of risk of bias for each included study. | No quantitative synthesis conducted |
| Results of individual studies | 19 | For all outcomes, present, for each study: (a) summary statistics for each group (where appropriate) and (b) an effect estimate and its precision (e.g. confidence/credible interval), ideally using structured tables or plots. | Tables 1 and 2, Figs 2-11 |
| Results of syntheses | 20a | For each synthesis, briefly summarise the characteristics and risk of bias among contributing studies. | No statistical synthesis conducted |
| | 20b | Present results of all statistical syntheses conducted. If meta-analysis was done, present for each the summary estimate and its precision (e.g. confidence/credible interval) and measures of statistical heterogeneity. If comparing groups, describe the direction of the effect. | No statistical synthesis conducted |
| | 20c | Present results of all investigations of possible causes of heterogeneity among study results. | No statistical synthesis conducted |
| | 20d | Present results of all sensitivity analyses conducted to assess the robustness of the synthesized results. | No statistical synthesis conducted |
| Reporting biases | 21 | Present assessments of risk of bias due to missing results (arising from reporting biases) for each synthesis assessed. | No statistical synthesis conducted |
| Certainty of evidence | 22 | Present assessments of certainty (or confidence) in the body of evidence for each outcome assessed. | No statistical synthesis conducted |
| **DISCUSSION** | | | |
| Discussion | 23a | Provide a general interpretation of the results in the context of other evidence. | Discussion section |

| Section and Topic | Item # | Checklist item | Location where item is reported |
|---|---|---|---|
| | 23b | Discuss any limitations of the evidence included in the review. | Methods, Discussion and Conclusion sections |
| | 23c | Discuss any limitations of the review processes used. | Methods, Discussion and Conclusion sections |
| | 23d | Discuss implications of the results for practice, policy, and future research. | Discussion and Conclusion sections |
| **OTHER INFORMATION** | | | |
| Registration and protocol | 24a | Provide registration information for the review, including register name and registration number, or state that the review was not registered. | the review was not registered. |
| | 24b | Indicate where the review protocol can be accessed, or state that a protocol was not prepared. | the review was not registered. |
| | 24c | Describe and explain any amendments to information provided at registration or in the protocol. | the review was not registered. |
| Support | 25 | Describe sources of financial or non-financial support for the review, and the role of the funders or sponsors in the review. | Information on funding sources provided to the editors of the journal |
| Competing interests | 26 | Declare any competing interests of review authors. | Information provided to the editors of the journal |
| Availability of data, code and other materials | 27 | Report which of the following are publicly available and where they can be found: template data collection forms; data extracted from included studies; data used for all analyses; analytic code; any other materials used in the review. | This will be made available upon publication |

**Supplementary Note S1. Boolean search terms**

The following Boolean search terms were used (in Web of Science syntax): *(((((TI=(manipulation OR disinformation OR misinformation OR propaganda OR "fake news" OR rumo\* OR "misleading information" OR "false information" OR "computational propaganda" OR conspira\* OR skepti\* OR scepti\*)) OR AB=(manipulation OR disinformation OR misinformation OR propaganda OR "fake news" OR rumo\* OR "misleading information" OR "false information" OR "computational propaganda" OR conspira\** OR skepti\* OR scepti\**) OR AK=(manipulation OR disinformation OR misinformation OR propaganda OR "fake news" OR rumo\* OR "misleading information" OR "false information" OR "computational propaganda" OR conspira\** OR skepti\* OR scepti\*))))) AND (TI=("social media" OR "social networking site" OR "platform\*") OR AB=("social media" OR "social networking site" OR "platform\*") OR AK=("social media" OR "social networking site" OR "platform\*" OR "social network")) AND (LA==("ENGLISH")) AND (TI=(climat\* OR "global warming") OR AB=(climat\* OR "global warming") OR AK=(climat\* OR "global warming"))*

**Supplementary Note S2. Eligibility criteria**

*Publication day*. Study was published between November 1, 2006 (after the public launch of Facebook, perhaps the first modern social media platform) and December 9, 2022 (date of data collection).

*English*. Study available in English.

*Manipulation*. Study discusses any aspects of online manipulation, including and referring to misinformation, disinformation, fake news, propaganda, and similar.

*Digital platforms*. Mention of any social media platform as a study object. We defined a social media platform as a website that is dependent on user-generated content to facilitate two-way interaction.

*Domain.* A study explores manipulation in relation to climate change.